\documentclass[
superscriptaddress,
reprint,
amsmath,amssymb,
aps,
pra
]{revtex4-2}
\usepackage[english]{babel}

\usepackage{caption}
\usepackage{subcaption}

\usepackage[colorlinks=false]{hyperref}

\usepackage{graphicx}% Include figure files
\usepackage{dcolumn}
\usepackage{bm}% bold math

\usepackage{hhline}
\usepackage{multirow}
\usepackage{dsfont}

\usepackage{physics}

\usepackage{float}
\usepackage{lipsum}

\begin{document}

    \title{Robustness of chaotic behavior in iterated quantum protocols
}

    \author{Attila Portik}
    \email{portik.attila@wigner.hun-ren.hu}
    \author{Orsolya K\'alm\'an}
    \affiliation{HUN-REN Wigner Research Centre for Physics, 1525 P.O. Box 49, Budapest, Hungary}
    \author{Igor Jex}
    \affiliation{Faculty of Nuclear Sciences and Physical Engineering, Czech Technical University in Prague, B\v rehov\'a 7, 115 19 Praha 1 - Star\'e M\v esto, Czech Republic}
    \author{Tam\'as Kiss}
    \affiliation{HUN-REN Wigner Research Centre for Physics, 1525 P.O. Box 49, Budapest, Hungary}
    \date{\today}
    
    \begin{abstract}
    One of the simplest possible quantum circuits, consisting of a CNOT gate, a Hadamard gate and a measurement on one of the outputs is known to lead to chaotic dynamics when applied iteratively on an ensemble of equally prepared qubits. The evolution of pure initial quantum states is characterized by a fractal (in the space of states), formed by the border of different convergence regions. We examine how the ideal evolution is distorted in the presence of both coherent error and incoherent initial noise, which are typical imperfections in current implementations of quantum computers. It is known that under the influence of initial noise only, the fractal is preserved, moreover, its dimension remains constant below a critical noise level. We systematically analyze the effect of coherent Hadamard gate errors by determining fixed points and cycles of the evolution. We combine analytic and numerical methods to explore to what extent the dynamics is altered by coherent errors in the presence of preparation noise as well. We show that the main features of the dynamics, and especially the fractal borders, are robust against the discussed noise, they will only be slightly distorted. We identify a range of error parameters, for which the characteristic properties of the dynamics are not significantly altered. Hence, our results allow to identify reliable regimes of operation of iterative protocols. 
    \end{abstract}
    
    \maketitle
    
    \section{Introduction}
    
    The emergence of efficient quantum protocols may result in a computational speedup over classical computation for a variety of tasks, such as cryptography \cite{365700}, simulating quantum systems \cite{1073} and machine learning \cite{nature23474}. The efficient realisation of a quantum protocol requires three essential steps to be carried out in an adequately precise manner: the preparation of quantum states, their coherent manipulation through the consecutive application of quantum gates, and their measurement. Current quantum processors are not yet capable of maintaining long enough coherence times to preserve the quantum information contained within quantum states due to decoherence. At the same time, they also lack the necessary size and complexity to allow for the use of advanced quantum error correction methods to reduce the effects of quantum noise \cite{Preskill2018quantumcomputingin}, hence they are termed Noisy Intermediate-Scale Quantum (NISQ) computers. In fact, in NISQ devices, any of the fundamental steps of a quantum protocol may be affected by noise. Faulty state preparation, imperfect quantum gates or measurements, and decoherence all may jeopardise the final result \cite{proctor2022measuring}. Consequently, the progress towards practical quantum computing is dependent upon the advancement of both quantum hardware and fault-tolerant quantum error correction algorithms. %To this end, an analysis of various forms of quantum decoherence and noise has recently been conducted in order to gain a better understanding of the problem \cite{0006204, SHUKLA2020126387}.
    
    The noise that occurs in quantum computers can arise from various sources, although it can be divided into two main types: systematic or coherent errors and stochastic or random noise. The random noise emerges from a stochastic source, such as environmental fluctuations of the parameters associated with the qubit system. This uncontrolled fluctuation leads to qubit decoherence. Coherent error, on the other hand, is usually the result of an underlying systematic error that occurs in the same form, such as miscalibrated quantum logic gates \cite{aa9a06, 042338}. Whenever the faulty quantum gate is applied, it distorts the state of the qubit more and more, so that eventually, after such a gate has been applied too many times, the original quantum information may be lost \cite{1.5089550, Ball2016}.
    
    The time evolution of closed quantum systems is deterministic and can be formulated in terms of linear maps of the density operator. However, if one measures a part of the system and post-selects according to a given measurement result, a protocol can be designed where the unobserved part of the system evolves according to a nonlinear transformation. Such a time evolution can be achieved by applying an entangling operation on two (or more) copies of a quantum system and then performing a measurement on some of the subsystems \cite{bechmann1998non, terno1999nonlinear, quant-ph/0008022} according to the results of which  one keeps or rejects the unmeasured part of the system. By repeating this procedure in an iterative manner on the kept systems, one can create an iterated nonlinear quantum protocol. Already the very basic protocol involving a CNOT gate and a subsequent single-qubit unitary operation on the post-selected state in each iterative step, can create a diverse range of different nonlinear protocols in the case of qubits \cite{kiss2006complex, kiss2011measurement}. Certain unitary operations lead to a complex deterministic chaotic time evolution for every pure initial state \cite{gilyen2016exponential, kalman2018sensitivity}. On the other hand, with the appropriate choice of the unitary transformation, unique nonlinear protocols can be constructed that are suitable for solving specific tasks, such as quantum state discrimination of qubits \cite{kalman2018quantum} or benchmarking quantum computers \cite{2210.09674, cornelissen2021scalable}. Moreover, by considering a similar scheme in a quantum communication scenario on entangled qubit pairs shared by two parties, one can construct entanglement distillation protocols \cite{guan2013reexamination, dur2007entanglement}. 
    
    In order to perform a large number of iterations of a nonlinear protocol, one needs to prepare and coherently manipulate a large number of qubits for long enough times. Therefore, the first physical implementations of these iterated protocols have so far been limited to a few steps, i.e., up to two iterations in optical experiments \cite{zhu2019experimental,qu2021observation}, and three in programmable quantum computers \cite{cornelissen2021scalable}. In the latter case, even though the available number of qubits is constantly increasing, the performance of the devices still needs to be improved in order to produce faithful results.

    Iterated nonlinear quantum protocols offer a promising way to study the effects of noise as they utilize the main building blocks of any quantum algorithm in an iterative manner. In the dynamics of certain protocols, a complicated fractal structure appears in the space of initial states, which separates different regions in which states converge to a specific outcome. The  properties of the fractal are determined by the actual unitary transformation applied in every step of the protocol. Small changes in the initial state do not result in a different outcome unless the initial state is located near the border of a domain. Assuming that the protocol is implemented in the presence of preparation noise, it is surprising to observe that the fractal is preserved up to a certain critical value of the noise \cite{malachov2019phase, viennot_competition_2022}. Above this critical level, the fundamental properties of the fractal do not change. Although the protocol itself can be particularly sensitive to the choice of the initial state, at the same time, it can tolerate errors which occur during state preparation. This feature of the protocol suggests that it may tolerate other types of incoherent errors as well \cite{nonunitary}. On the other hand, its iterative structure might render it susceptible to coherent errors stemming from the application of faulty quantum logic gates, since the repeated application of the same quantum logic gates could amplify the effect of any coherent, systematic errors and thus represent a significant source of noise.
    
    In this paper we study a quadratic iterated quantum protocol involving a CNOT gate and a subsequent Hadamard gate on the kept qubit in every step. We assume that the CNOT gate is ideal, but the Hadamard gate is affected by coherent errors. We investigate how the characteristics of the protocol, specifically, the corresponding fractal structure, changes. We show that for small coherent errors the most significant features of the protocol are very similar to the case when no coherent errors are present. In order to further investigate these similarities, we also study such "faulty" protocols in the presence of state preparation noise and conclude that these two types of errors can both be tolerated by the scheme up to a certain amount.
    
    The paper is organized as follows. In Section \ref{Characteristics of the ideal protocol}, we introduce the iterated nonlinear quantum protocol and its characteristic properties for pure initial states as well as mixed ones (accounting for state preparation noise). In Sec. \ref{Coherent single-qubit gate errors}, we propose a model for the faulty Hadamard gate, where we assume that the coherent error comes from the miscalibrated single-qubit gates that realize the Hadamard gate. In Sec. \ref{Effect of coherent error}, we show how the characteristics of the protocol change when coherent gate errors occur and when both preparation noise and coherent errors appear simultaneously. Section \ref{Phase transition in presence of coherent gate errors} focuses on the variation of the characteristic phase transition as the consequence of the systematic error of the miscalibrated gate. Finally, in Sec. \ref{Small coherent error}, we identify an error limit beyond which the essential characteristics of the protocol do not change significantly compared to the error-free case. We conclude in Sec.~\ref{Discussion}.

    \section{Characteristics of the ideal protocol} \label{Characteristics of the ideal protocol}
    
\begin{figure}[H]
        \begin{subfigure}{0.44\columnwidth}
            \includegraphics[width=\linewidth]{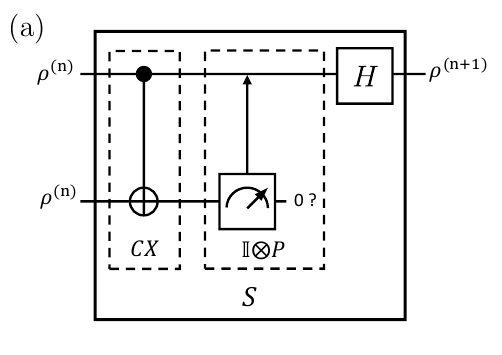}
            \phantomsubcaption
            \label{fig : 1_a}
        \end{subfigure}
        \begin{subfigure}{0.52\columnwidth}
            \includegraphics[width=\linewidth]{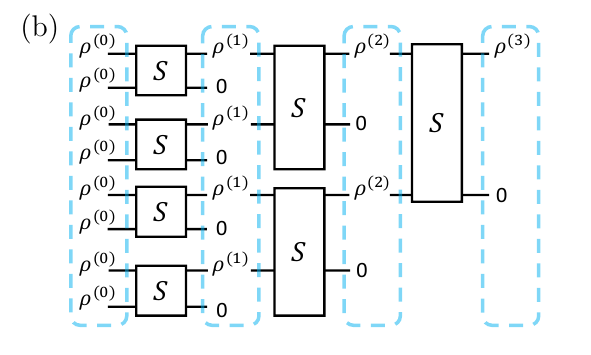}
            \phantomsubcaption
            \label{fig : 1_b}
        \end{subfigure}
        \caption{(a) Quantum circuit realizing a single step ($S$) of the protocol: A CNOT gate acts on two copies of the quantum state, then the target qubit is measured in the computational basis. If the outcome of the measurement is $0$, then a Hadamard gate is applied to the other, unmeasured qubit. (b) Schematic quantum circuit implementing three iterations of the protocol.}
        \label{fig : 1}
    \end{figure}

    In this work, we focus on the effect of coherent gate errors on a previously studied iterated nonlinear quantum protocol \cite{malachov2019phase}. A single step of the protocol requires a pair of qubits in the same initial quantum state $\ket{\psi_0}$ as inputs of a CNOT gate. After the CNOT, one measures the target qubit, and keeps the control qubit only if the target was measured to be $\ket{0}$, and then applies a Hadamard gate on the kept qubit (see Fig.~ \ref{fig : 1_a}). By using the Riemann parameterization of single-qubit states the input state can be written as $\ket{\psi_0} =  N_0 \left(\ket{0} + z \ket{1} \right) \,$, where $z \in \mathbb{C}_{\infty}$, and $N_0=\frac{1}{\sqrt{1+|z|^2}}$. It can be shown that the procedure leads to an output state that can be written as $\ket{\psi_1}=N_1 \left(\ket{0} + f(z) \ket{1} \right)$, where $f(z)=\frac{1-z^2}{1+z^2}$ is a complex quadratic rational function. Thus, the input state is transformed nonlinearly in a single step of the protocol. 
    
\begin{figure}[H]
        \begin{subfigure}{0.5\columnwidth}
            \includegraphics[width=\linewidth]{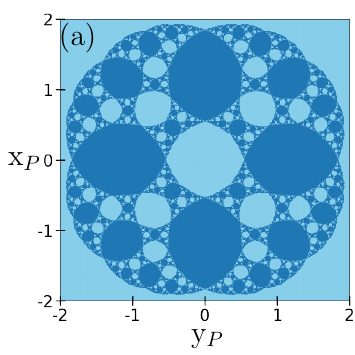}
            \phantomsubcaption
            \label{fig : 2_a}
        \end{subfigure}%
        \begin{subfigure}{0.5\columnwidth}
            \includegraphics[width=\linewidth]{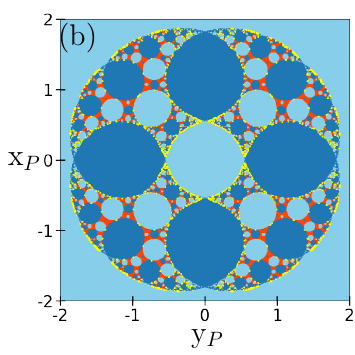}
            \phantomsubcaption
            \label{fig : 2_b}
        \end{subfigure}%
        
        \begin{subfigure}{0.5\columnwidth}
            \includegraphics[width=\linewidth]{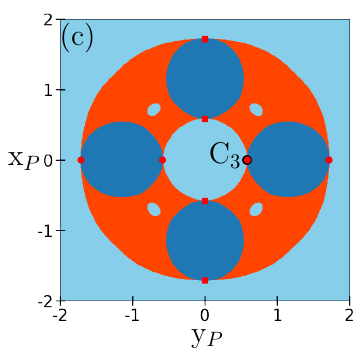}
            \phantomsubcaption
            \label{fig : 2_c}
        \end{subfigure}%
        \begin{subfigure}{0.5\columnwidth}
            \includegraphics[width=\linewidth]{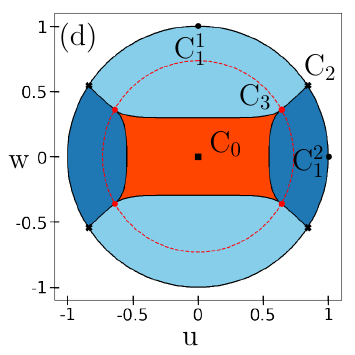}
            \phantomsubcaption
            \label{fig : 2_d}
        \end{subfigure}%
        
        \caption{(a)-(c) Stereographic projections of spherical surfaces corresponding to initial purity values $P = 1$, $0.95$, and $0.75$, respectively. Initial states represented with light (dark) blue color converge to the point $C_1^1$ ($C_1^2$) after an even number of iterations. Red color marks initial states which converge to the maximally mixed state ($C_0$). Yellow dots in (b) mark the points of the backwards-iterated Julia set. Note that the stereographic projection is taken from the south pole of the spherical surface onto the $w=0$ plane, so that $x_P+iy_P = \frac{u + i v }{\sqrt{2P -1} +w}$. (d) The invariant plane $(u,0,w)$ of the map of Eq.~(\ref{mixed_map}). Coloring is the same as in (a). }
        \label{figure : 2} 
    \end{figure}

    Provided we have multiple copies of the same input state $\ket{\psi_0}$, we can construct quantum circuits where we combine two such nonlinearly transformed qubits in state $\ket{\psi_1}$ as inputs of a similar second step, which, when succeeding, produces an output state $\ket{\psi_2}=N_2(\ket{0}+f^{(2)}(z)\ket{1}$, where $f^{(2)}=f(f(z))$ is the second iterate of the complex quadratic rational function $f(z)$. One can then continue in the same fashion for yet another step, and so on as shown in Fig.~\ref{fig : 1_b}. In this way, $n$ successful subsequent steps of the protocol will lead to an output state of the form $\ket{\psi_n}=N_n(\ket{0}+f^{(n)}(z)\ket{1})$, where $f^{(n)}(z)$ is the $n$th iterate of $f(z)$. 
    As the procedure assigns single-qubit states to single-qubit states which can be described by the iterates of the complex rational function $f(z)$, the qubit dynamics can be analyzed by studying the characteristics of $f(z)$, such as repelling points, attractors and their associated basins of attraction, etc. It can be easily shown that $f(z)$ has a single attractive cycle of length two: $z_1^1=0 \leftrightarrow z_1^2=1$. The basins of attraction of these two points (they form together the so-called Fatou set) are separated by a fractal (the so-called Julia set) on the complex plane (see Fig.~\ref{fig : 2_a}). Both sets are invariant under $f$: the Fatou set contains the 'regular' points, while the Julia set contains all repelling cycles and corresponds to chaotic dynamics \cite{milnor2011dynamics}. 
    
    The above mentioned mathematical analysis of the pure-state dynamics can only describe an ideal, noiseless scenario. Current quantum processors, however, are still not advanced enough to achieve fault tolerance, therefore the effect of noise must also be taken into account in the model. In Ref.~\cite{malachov2019phase} the authors assumed preparation noise to be present in the system (without any quantum gate errors). The initial states can then be described by density operators, and the nonlinear transformation corresponding to a single step of the iterated protocol is given by
    \begin{equation}
        \rho^{(n)} \rightarrow \rho^{(n+1)} = H \dfrac{\rho^{(n)} \odot \rho^{(n)} }{\Tr \left( \rho^{(n)} \odot \rho^{(n)} \right)} H^{\dagger}
        \label{transf_rho}
    \end{equation}
    where $\rho^{(n)}$ is the input, while $\rho^{(n+1)}$ is the output density matrix of the $(n+1)$th step of the protocol, and the $\odot$ symbol stands for elementwise product in the computational basis (also known as Hadamard product).
    
    In this description, it is practical to parameterize the density matrix as
    \begin{equation}
        \rho = \dfrac{1}{2} \left[ 
        \begin{array}{ll}
            1 + w  & u-iv \\
            u + iv & 1 - w  
        \end{array}\right] \, ,
    \end{equation}
    where $u,v,w \in \mathbb{R}$ are the Bloch-sphere coordinates of $\rho$ and the purity $P=\mathrm{Tr}(\rho^2)=(1+u^2+v^2+w^2)/2\leq1$ can be used to quantify the noisiness of $\rho$. With these three real parameters, the transformation in Eq.~(\ref{transf_rho}) can be converted to the following $\mathbb{R}^3 \rightarrow \mathbb{R}^3$ nonlinear map
    \begin{align}
        u_{n+1} & =  \dfrac{2w_n}{1+w_n^2} \, , \nonumber \\
        v_{n+1} & =  \dfrac{-2 u_n v_n}{1+w_n^2} \, , \label{mixed_map} \\
        w_{n+1} & =  \dfrac{u_n^2-v_n^2}{1+w_n^2} \,.  \nonumber
    \end{align}
    The dynamics of the system can be analyzed by studying the iterative properties of this map. 
    
    The analytical and numerical calculations in Ref.~\cite{malachov2019phase} revealed that the map %of Eq.~(\ref{mixed_map}) 
    has three attractive periodic points: a length-2 pure cycle consisting of points $C_1^1=(0,0,1)$ and $C_1^2=(1,0,0)$ (note that this cycle is the same as the one found in the case of the pure-state dynamics). The third attractive point is the maximally mixed state $C_0=(0,0,0)$. As a consequence, some initial states purify into the cycle $C_1$, some converge to the maximally mixed state $C_0$, and there are also states which behave chaotically, or quasi-chaotically (we will discuss this latter phenomenon in more detail below).
    
    An intriguing feature of the dynamics is that the $v=0$ plane, which corresponds to real density matrices, is an invariant subset of the map. Its identification greatly facilitated the numerical analysis in Ref.~\cite{malachov2019phase} as it contains the attractive points $C_1^1$, $C_1^2$ and $C_0$, as well as some repelling points of the pure-state dynamics: $C_2$ and its pre-images, which are members of the Julia set of $f$ (see Fig.~\ref{fig : 2_d}). There is also a repelling mixed fixed point ($C_3$) located on this plane that plays a special role in the noisy dynamics. 
    
    The asymptotic dynamics of initial states with a given purity can be visualized with the aid of the stereographic projection of the corresponding spherical surface of the Bloch sphere to the complex plane, as shown in Figs.~\ref{fig : 2_b} and \ref{fig : 2_c}. 
    Surprisingly, the fractal structure is not destroyed immediately with the addition of noise. Even though more and more states converge to the maximally mixed state as the initial purity is decreased, a fractal structure remains present up to a certain critical purity $P_c$. Moreover, the fractal dimension is constant above this critical purity. Reaching $P_c$ however, the fractal disappears and the fractal dimension drops to $1$ (see Fig.~\ref{fig : 9_a}). This phenomenon has the character of a phase transition, where the purity of the initial states is the control parameter.
    
    The constant fractal dimension 
    suggests that the protocol is, in some sense, robust against preparation noise. The fact that the complexity of the convergence pattern remains unchanged regardless of the presence of noise, indicates that the information content is retained down to the critical purity. 
    
    \section{Coherent single-qubit gate errors} \label{Coherent single-qubit gate errors}
    
    Random errors usually decrease the purity of a quantum state. However, it is plausible to assume that the above protocol may be able to correct these random errors in a similar way as it compensates for the effect of preparation noise. In the case of preparation error, if the quantum state of the qubit is displaced from the ideal, but the random deviation is appropriately small or the given state is not at the brink of the basins of attraction, then the result of the protocol will remain unchanged since the state remains in the same attraction region as it was in the ideal case. On the other hand, coherent errors, which arise systematically due to, for example, miscalibrated quantum gates, may lead to significant deviations from the ideal case, as in an iterated protocol, the same coherent error occurs each time the faulty quantum gate is used.

\begin{figure}[H] 
        \centering
        \includegraphics[width=\linewidth]{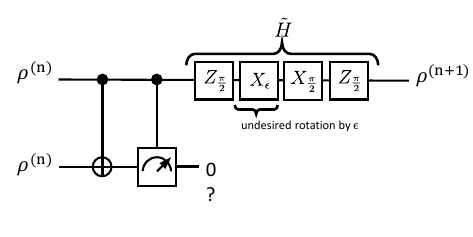}
        \caption{The decomposition of the faulty Hadamard gate in the quantum circuit that realises one step of the nonlinear protocol. The coherent gate error appears as an undesired $X_{\epsilon}$ rotation.}
        \label{fig : 3} 
    \end{figure}

    In what follows, we include the coherent error of the Hadamard gate in the model of the nonlinear protocol discussed in the previous section. In order to do so, we decompose the Hadamard gate into three consecutive rotations 
    along the $x$ and $z$ axes of the Bloch sphere as
    \begin{equation}
        \renewcommand*{\arraystretch}{1.2}
        H \equiv Z_{\pi/2} X_{\pi/2} Z_{\pi/2} 
        = - \dfrac{\sqrt{2}i}{2} 
        \left[
        \begin{matrix}
            1 & 1 \\
            1 & -1 
        \end{matrix}
        \right] \, .
    \end{equation}
    This decomposition contains two different rotations only. Moreover, in certain quantum computers (e.g., superconducting ones), the $Z$ gates can be implemented virtually -- via changing the frame of reference -- with zero error and duration \cite{mckay_efficient_2017-1, Johnson2015}. Thus, these so-called virtual $Z$ gates can be considered practically error-free, and one can assume that the coherent error of the $H$ gate, if present, mainly originates from the miscalibration of the $X_{\pi/2}$ gate. 
    The miscalibration of a single-qubit rotation can be interpreted as a small undesired extra rotation  $\epsilon$. In our case, this means that instead of $H$, a slightly different $\tilde{H}$ Hadamard gate is applied in every step, which can be written as
    \begin{align}
        \tilde{H}  & = Z_{\pi/2} X_{\pi/2} X_{\epsilon} Z_{\pi/2} \nonumber \\
                   & = e^{i \phi} \left[
        \begin{array}{lr}
            \cos{\left(\frac{\epsilon}{2} + \frac{\pi}{4} \right)} & \sin{\left(\frac{\epsilon}{2} + \frac{\pi}{4} \right)}\\
            \sin{\left(\frac{\epsilon}{2} + \frac{\pi}{4} \right)} & - \cos{\left(\frac{\epsilon}{2} + \frac{\pi}{4} \right)}
        \end{array}
        \right] \, ,
        \label{Eq : 5}
    \end{align}
    where the $X_{\epsilon}$ term accounts for the errors resulting from miscalibration (see Fig. \ref{fig : 3}).
    
    Assuming there are no other sources of errors, the dynamics of the protocol can still be described by a complex quadratic rational function, which can be written as 
    \begin{equation}
        f_{\epsilon}\left(z\right) = \dfrac{\sin{\left(\dfrac{\pi}{4} + \dfrac{\epsilon}{2} \right)} - z^{2} \cos{\left(\dfrac{\pi}{4} + \dfrac{\epsilon}{2} \right)}}{\cos{\left(\dfrac{\pi}{4} + \dfrac{\epsilon}{2}\right)} + z^{2} \sin{\left(\dfrac{\pi}{4} + \dfrac{\epsilon}{2} \right)}} \, 
    \end{equation}
    mapping pure states to pure states. (Note that for $\epsilon=0$ $f_{\epsilon}=f$.) 
    
    Similarly, one can determine the $\mathbb{R}^3 \rightarrow \mathbb{R}^3$ nonlinear map describing the time evolution of the system in the presence of both preparation noise and coherent errors (using the method presented in \cite{portik_iterated_2022}):
    \begin{align}
        \label{eq : time_evoltion}
        u_{k+1} & = \dfrac{2w_k \cos{\left(\epsilon \right)} + \left(u_k^{2} - v_k^{2}\right) \sin{\left(\epsilon \right)}}{1 + w_k^{2}} \, ,\nonumber \\  
        v_{k+1} & = \dfrac{- 2 u_k v_k}{1 + w_k^{2}} \, , \\ 
        w_{k+1} & = \dfrac{ \left(u_k^{2} - v_k^{2}\right) \cos{\left(\epsilon \right)}- 2 w_k \sin{\left(\epsilon \right)}}{1 + w_k^{2}} \, .\nonumber 
    \end{align}
    
    We will analyze the characteristic properties of these maps in the following sections.
    
    \section{Effect of coherent error} \label{Effect of coherent error}
    
    We examine the effect of the emerging coherent gate error through investigating the basic properties of the dynamical system associated with Eq.~(\ref{eq : time_evoltion}). By analyzing the characteristics of this system, such as the invariant sets, the repelling and attractive points and their basins of attraction, or the structure of the set of chaotic states, we can reveal which properties of the protocol are significantly changed compared to the error-free case.
    
    The analysis of Eq.~(\ref{eq : time_evoltion}) reveals that the dynamical system has two invariant sets: the surface of the Bloch sphere (i.e., the set of pure states) and, similarly to the coherent-error-free case, the $v=0$ plane. 
    
    \subsection{Properties of the invariant sets} \label{Properties of the invariant sets}
    
    There are three fixed points among the pure states, which can be determined by solving the polynomial equation
    \begin{equation}
        z^3 + \cot{\left(\dfrac{\epsilon}{2} + \frac{\pi}{4} \right)}z^2 + \cot{\left(\dfrac{\epsilon}{2} + \frac{\pi}{4} \right)} z - 1 = 0  \, .
    \end{equation}
    One of them is always real, which means that it is located at the intersection of the two invariant sets. By calculating the multiplier (i.e., the absolute value of the derivative of the function $f_{\epsilon}$) at these points reveals that all of them are repelling.
    
\begin{figure}[H]
        \begin{subfigure}{0.45\columnwidth}
            \includegraphics[width=\linewidth]{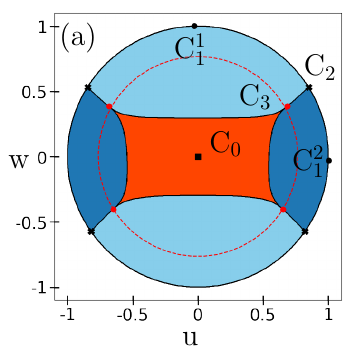}
            \phantomsubcaption
            \label{fig : 4_a}
        \end{subfigure}
        \begin{subfigure}{0.45\columnwidth}
            \includegraphics[width=\linewidth]{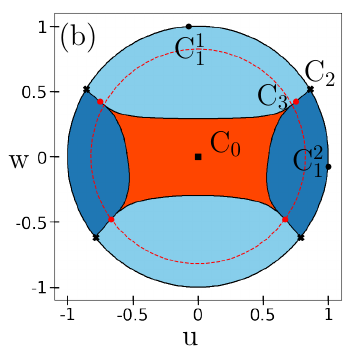}
            \phantomsubcaption
            \label{fig : 4_b}
        \end{subfigure}
        
        \begin{subfigure}{0.45\columnwidth}
            \includegraphics[width=\linewidth]{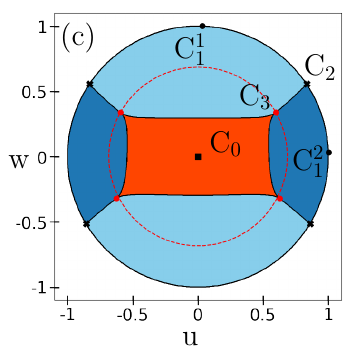}
            \phantomsubcaption
            \label{fig : 4_c}
        \end{subfigure}
        \begin{subfigure}{0.45\columnwidth}
            \includegraphics[width=\linewidth]{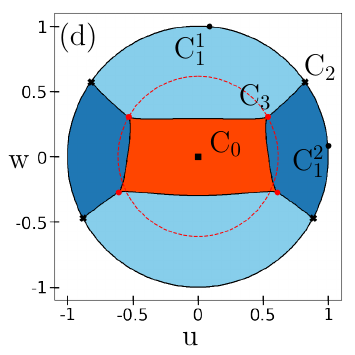}
            \phantomsubcaption
            \label{fig : 4_d}
        \end{subfigure}
        
        \caption{The convergence regions of the nonlinear map $f_{\epsilon}$ on the $(u,0,w)$ invariant plane for different values of $\epsilon$. (a)-(d) figures correspond to $\epsilon=1.8^\circ \left( 2 \% \right)$, $4.5^\circ \left( 5\% \right)$, $-1.8^\circ \left( -2\% \right)$ and $-4.5^\circ \left( -5\% \right)$, respectively. Initial states represented with light (dark) blue converge to the point $C_1^1$ ($C_1^2$) of pure limit cycle $C_1$ after an even number of steps. Red colour indicates states that converge to the maximally mixed state ($C_0$). The red dashed line represents the purity of the least pure preimage of the repelling fixed point $C_3$.}
        \label{fig : 4} 
    \end{figure}

\begin{figure}[H]
        \centering
        \includegraphics[width=\columnwidth]{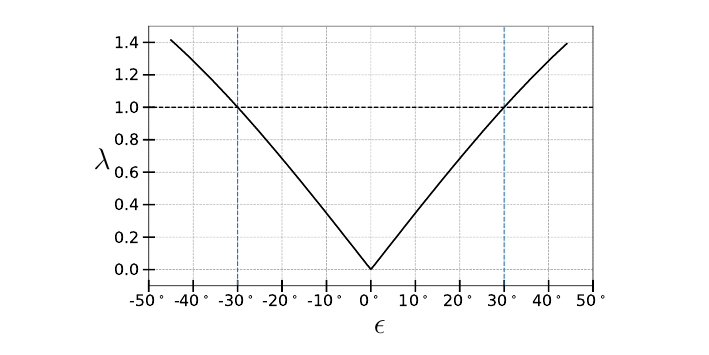}
        \caption{The multiplier $\lambda$ of the maximally mixed state $C_0$ as a function of $\epsilon$. $0<\lambda<1$ corresponds to $C_0$ being attractive, while $\lambda>1$ means that $C_0$ is repelling.}
        \label{fig : 12}
    \end{figure}

\begin{figure} 
        \centering
        \includegraphics[width=\columnwidth]{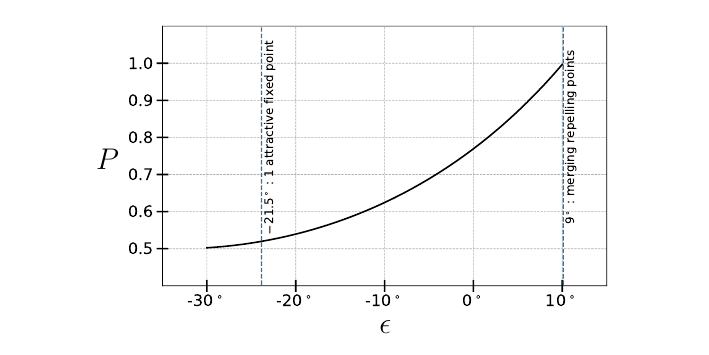}
        \caption{The purity of the repelling mixed fixed point $C_3$ as a function of $\epsilon$.}
        \label{fig : 11}
    \end{figure}

    Fixed points, located on the $v=0$ invariant plane of the dynamics can be calculated by finding the roots of the coupled equations
    \begin{align}
    \label{eq : v=0 fixed points}
        u & = \dfrac{2w \cos{\left(\epsilon \right)} + u^{2} \sin{\left(\epsilon \right)}}{1 + w^{2}},  \\ 
        w & = \dfrac{u^{2} \cos{\left(\epsilon \right)}- 2 w \sin{\left(\epsilon \right)}}{1 + w^{2}}. \nonumber
    \end{align}
    Similarly to the $\epsilon = 0$ case, there exist two mixed fixed points on the invariant plane for any value of $\epsilon$. One of these points is the maximally mixed state $C_0$, irrespective of the value of $\epsilon$, the other one is a mixed fixed point, whose coordinates are $\epsilon$-dependent: Changing $\epsilon$ continuously, the fixed point shifts along the boundary of the different regions of attraction on the invariant plane (see Fig.~\ref{fig : 4}), and in the $\epsilon\rightarrow 0$ limit it coincides with the mixed fixed point $C_3$ of the error-free case. By evaluating the spectral radius of the Jacobi matrix at these fixed points (which we will also denote by $C_3$), it can be shown that they are always repelling, as shown in Table~\ref{table : critical fixed point}. The value of the spectral radius for the $C_0$ maximally mixed state reveals that it is attractive for a relatively small magnitude of the coherent error (see Fig.~\ref{fig : 12}). As the amount of undesired rotation is increased, the range of attraction of the maximally mixed state becomes smaller and smaller. At the same time, the repelling point $C_3$ approaches the centre of the Bloch sphere (see Fig.~\ref{fig : 11}). Around $\left| \epsilon \right| =30^{\circ}$, these two points with opposite properties get so close to each other, that the maximally mixed state is overshadowed by $C_3$ and it is no longer attractive (see Figs.~\ref{fig : 11} and \ref{fig : 12}).
    
    As it can be seen in Fig.~\ref{fig : 11}, in the case of over-rotation the purity of $C_3$ increases continuously and even accelerates until it reaches the set of pure states at $\epsilon = 10^\circ$. From that point on, $C_3$ and the pure repelling fixed point $C_2$ are no longer distinguishable. For larger values of $\epsilon$ there is no mixed repelling fixed point in the dynamics.   
    
    \begin{table}[H]
    \centering
        \begin{tabular}{cccc} \hline
            $\epsilon [^{\circ}]$ & $C_3$ & $P_3$ & $\lambda$ \\ 
            \hline \hline
            $10$  & $(0.877, 0, 0.48)$  & $ 1 $     & $ 1.426 $ \\ 
            $9$   & $(0.871, 0, 0.484)$ & $ 0.997 $ & $ 1.532 $ \\ 
            $8$   & $(0.844, 0, 0.47)$  & $ 0.966 $ & $ 1.417 $ \\ 
            $7$   & $(0.817, 0, 0.456)$ & $ 0.937 $ & $ 1.352 $ \\ 
            $6$   & $(0.79, 0, 0.442)$  & $ 0.91 $  & $ 1.451 $ \\ 
            $5$   & $(0.764, 0, 0.428)$ & $ 0.883 $ & $ 1.342 $ \\ 
            $4$   & $(0.738, 0, 0.414)$ & $ 0.858 $ & $ 1.323 $ \\ 
            $3$   & $(0.713, 0, 0.401)$ & $ 0.834 $ & $ 1.394 $ \\ 
            $2$   & $(0.688, 0, 0.388)$ & $ 0.812 $ & $ 1.257 $ \\ 
            $1$   & $(0.663, 0, 0.374)$ & $ 0.79 $  & $ 1.384 $ \\ 
            $0$   & $(0.639, 0, 0.361)$ & $ 0.769 $ & $ 1.361 $ \\ 
            $-1$  & $(0.615, 0, 0.348)$ & $ 0.75 $  & $ 1.311 $ \\ 
            $-2$  & $(0.591, 0, 0.335)$ & $ 0.731 $ & $ 1.461 $ \\ 
            $-3$  & $(0.568, 0, 0.322)$ & $ 0.713 $ & $ 1.351 $ \\ 
            $-4$  & $(0.545, 0, 0.31)$  & $ 0.696 $ & $ 1.375 $ \\ 
            $-5$  & $(0.522, 0, 0.297)$ & $ 0.68 $  & $ 1.549 $ \\ 
            $-6$  & $(0.499, 0, 0.285)$ & $ 0.665 $ & $ 1.364 $ \\ 
            $-7$  & $(0.477, 0, 0.272)$ & $ 0.651 $ & $ 1.405 $ \\ 
            $-8$  & $(0.455, 0, 0.26)$  & $ 0.637 $ & $ 1.64 $  \\ 
            $-9$  & $(0.433, 0, 0.248)$ & $ 0.624 $ & $ 1.399 $ \\ 
            $-10$ & $(0.411, 0, 0.235)$ & $ 0.612 $ & $ 1.39 $  \\   \hline
        \end{tabular}
        \caption{The coordinates and the multiplier of the internal repelling fixed point $C_3$ for different values of the error parameter $\epsilon$.}
        \label{table : critical fixed point}
    \end{table}

    In Sec.~\ref{Characteristics of the ideal protocol} we have seen that in the ideal ($\epsilon=0$) case, there was an attractive length-2 cycle in the dynamics. It is plausible to assume that, up to a certain value of $\epsilon$, similarly to the above mentioned fixed points, a length-2 cycle exists as well. This can be determined by analytically solving the equation $f_{\epsilon}\left(f_{\epsilon}\left(z\right)\right) = z$. The resulting periodic trajectory for a given $\epsilon$ can be shown to be attractive by determining the corresponding multiplier $\lambda$, as shown in Table \ref{table : pure attractive points}. (Note that we use the same notation for these points, namely $C_1^1$ and $C_1^2$, as in the $\epsilon=0$ case.) It can be seen that while in the $\epsilon=0$ case the length-2 cycle is superattractive ($\lambda=0$), as $\epsilon$ is increased in either direction, the attractiveness of the cycle decreases ($\lambda>0$). The points of the cycle are shifted towards (away from) each other when $\epsilon>0$ ($\epsilon<0$), see Fig.~\ref{fig : 4}. 
    
    \begin{table}
        \centering
        \begin{tabular}{ccc} \hline
            $\epsilon [^\circ]$ & $C_1^1 \leftrightarrow C_1^2$ & $\lambda$  \\ 
            \hline \hline 
            $ 10 $  & $(-0.148,0,0.989) \leftrightarrow (0.987,0,0.161)$    & $0.287$   \\ 
            $ 9 $   & $(-0.135,0,0.991) \leftrightarrow (0.989,0,0.148)$    & $0.264$   \\ 
            $ 8 $   & $(-0.122,0,0.993) \leftrightarrow (0.991,0,0.134)$    & $0.24$    \\ 
            $ 7 $   & $(-0.109,0,0.994) \leftrightarrow (0.993,0,0.118)$    & $0.214$   \\ 
            $ 6 $   & $(-0.095,0,0.995) \leftrightarrow (0.995,0,0.1)$      & $0.187$   \\ 
            $ 5 $   & $(-0.08,0,0.997) \leftrightarrow (0.996,0,0.089)$     & $0.159$   \\ 
            $ 4 $   & $(-0.065,0,0.998) \leftrightarrow (0.998,0,0.063)$    & $0.13$    \\ 
            $ 3 $   & $(-0.05,0,0.999) \leftrightarrow (0.999,0,0.045)$     & $0.099$   \\ 
            $ 2 $   & $(-0.034,0,0.999) \leftrightarrow (0.999,0,0.045)$    & $0.067$   \\ 
            $ 1 $   & $(-0.017,0,1.0) \leftrightarrow (1.0,0,0.0)$          & $0.034$   \\ 
            $ 0 $   & $(0,0,1) \leftrightarrow (1,0,0)$                     & $0$       \\ 
            $ -1 $  & $(0.018,0,0.999) \leftrightarrow (1,0,0)$             & $0.036$   \\ 
            $ -2 $  & $(0.036,0,0.999) \leftrightarrow (0.999,0,0.045)$     & $0.072$   \\ 
            $ -3 $  & $(0.055,0,0.998) \leftrightarrow (0.999,0,0.045)$     & $0.11$    \\ 
            $ -4 $  & $(0.075,0,0.997) \leftrightarrow (0.997,0,0.077)$     & $0.149$   \\ 
            $ -5 $  & $(0.096,0,0.995) \leftrightarrow (0.996,0,0.089)$     & $0.19$    \\ 
            $ -6 $  & $(0.117,0,0.993) \leftrightarrow (0.994,0,0.109)$     & $0.231$   \\ 
            $ -7 $  & $(0.139,0,0.99) \leftrightarrow (0.991,0,0.134)$      & $0.273$   \\ 
            $ -8 $  & $(0.162,0,0.987) \leftrightarrow (0.988,0,0.154)$     & $0.317$   \\ 
            $ -9 $  & $(0.186,0,0.983) \leftrightarrow (0.985,0,0.173)$     & $0.362$   \\ 
            $ -10 $ & $(0.211,0,0.977) \leftrightarrow (0.981,0,0.194)$     & $0.408$   \\  \hline
        \end{tabular}
        \caption{The coordinates of the attractive pure states on the Bloch sphere ($C_1^2 \leftrightarrow C_1^2$) for different values of the error parameter ($\epsilon$). The corresponding multipliers ($\lambda$) reveal that all of these cycles are attractive.}
        \label{table : pure attractive points}
    \end{table}

    Let us point out here that it is known from the theory of complex dynamical systems that the number of attractive periodic points of a complex function cannot exceed the number of different critical points of the function, since every direct attraction region must contain at least one critical point \cite{milnor2011dynamics}. The critical points of a complex function can be determined from the equation $f'_{\epsilon}\left(z\right)=0$. It is easy to see that for any value of $\epsilon$, $f_{\epsilon}$ has only two critical points: $0$ and $\infty$. Thence, $f_{\epsilon}$ cannot have more than two pure attractive periodic points, and since we found a length-2 attractive cycle for every $\epsilon$ in Table~\ref{table : pure attractive points}, we have determined all the pure, stable states. 
    
    Studying the dynamics on the invariant plane as shown in Fig.~\ref{fig : 4}, one can see that the protocol, apart from cases of large coherent error, can result in three different stable outcomes, which are very similar to the coherent-error-free case (see Fig.~\ref{fig : 2_d}): If the preparation noise is not too large, then the protocol purifies the qubit state into one of the pure, stable states of cycle $C_1$, while for larger preparation noise, the qubit states converge to the maximally mixed state. As a consequence of the coherent error, the points of  cycle $C_1$ are slightly shifted compared to the noise-free case (see also Table \ref{table : pure attractive points}). The regions of attraction on the invariant plane exhibit a similar arrangement regardless of the magnitude of the undesired rotation; however, their exact shape and position vary as the error parameter changes. The extent and direction of these distortions depend on the magnitude and sign of $\epsilon$. In the case of under-rotation (Figs.~\ref{fig : 4_c} and \ref{fig : 4_d}), parts of the attraction region in the northern hemisphere shrink, while in the southern hemisphere they expand. For over-rotations (Figs.~\ref{fig : 4_a} and \ref{fig : 4_b}), this behavior is just the opposite.  At the same time, the region of attraction of the maximally mixed state is continuously getting closer to the surface for larger and larger values of $\epsilon$. The position of the repelling mixed fixed point $C_3$ is shifted as $\epsilon$ is varied, but it always lies in the positive quadrant, at the point where the three different basins of attraction touch. Let us point out that in the $\epsilon=0$ case, it was shown that there are similar points in the other three quadrants which are in fact preimages of $C_3$, having the same purity. In the $\epsilon\neq 0$ case, as the shape of the different convergence regions is no longer symmetric to the $w=0$ axis, the purity of two of these points is different. 
    
\begin{figure}
        \begin{center}
            \begin{subfigure}{0.45\columnwidth}
                \includegraphics[width=\linewidth]{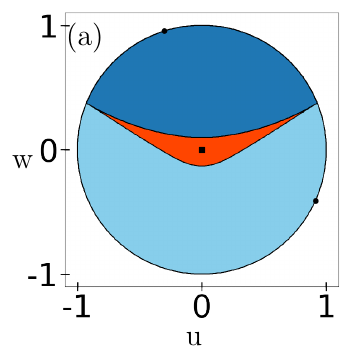}
                \phantomsubcaption
                \label{fig : 13_a}
            \end{subfigure}
            \begin{subfigure}{0.45\columnwidth}
                \includegraphics[width=\linewidth]{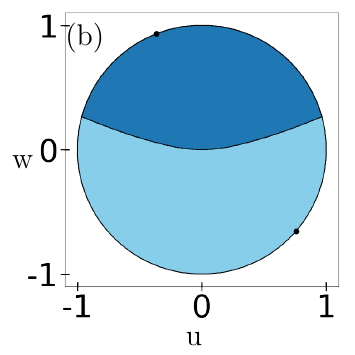}
                \phantomsubcaption
                \label{fig : 13_b}
            \end{subfigure}
            
            \begin{subfigure}{0.45\columnwidth}
                \includegraphics[width=\linewidth]{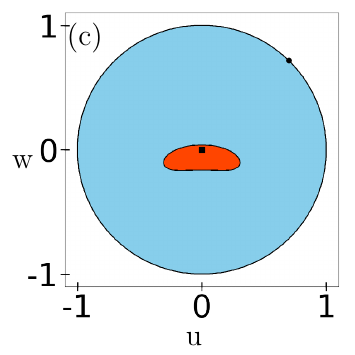}
                \phantomsubcaption
                \label{fig : 13_c}
            \end{subfigure}
            \begin{subfigure}{0.45\columnwidth}
                \includegraphics[width=\linewidth]{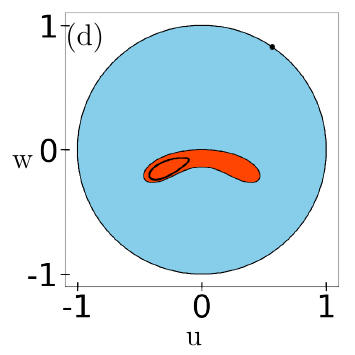}
                \phantomsubcaption
                \label{fig : 13_d}
            \end{subfigure}
            
            \caption{The regions of attraction on the $v=0$ invariant plane. Figs. (a)-(d) correspond to $\epsilon=27^\circ \left(30\%\right)$, $45^\circ \left(50\%\right)$, $-27^\circ \left(-30\%\right)$ and $-45^\circ \left(-50\%\right)$, respectively. Coloring is the same as in Fig.~\ref{fig : 4}, with the exception of (d), where the red color represents convegence to a long attractive mixed cycle, denoted by the black points inside the red region.}
            \label{fig : 13} 
        \end{center}
    \end{figure}

    One can observe that in the case of large enough over-rotations ($\epsilon\geq 27^{\circ}$), the basin of attraction of the maximally mixed state reaches the immediate surrounding of the pure-state surface on the invariant plane (see Fig.~\ref{fig : 13_a}). Thus, there exist points on the invariant plane where an arbitrarily small amount of preparation error can hinder the purification of the state. If the angle of over-rotation is even larger ($\epsilon> 30^\circ$), then the maximally mixed state loses its attractiveness and becomes repelling (see Fig. \ref{fig : 13_b}). In these cases, there is no attractive point inside the Bloch sphere any more.
    
    When the faulty Hadamard gate suffers from an under-rotation ($\epsilon<0^{\circ}$), the elements of the pure attractive cycle gradually approach each other for smaller and smaller values of $\epsilon$. At $\epsilon = -21.5^\circ$, the two points merge, and below this value, only a single pure attractive fixed point exists. The maximally mixed state remains attractive, and the repelling fixed point $C_3$ is situated at the boundary of the two basins of attraction, i.e., that of the sinlge pure attractor and the maximally mixed state (see Fig. \ref{fig : 13_c} and Table \ref{table : pure attractive points 2}). If the angle of under-rotation is less than $\epsilon = -30^\circ$, where the maximally mixed state becomes repelling, a new attractive mixed cycle appears and practically takes over the role of the $C_0$ point (see Fig. \ref{fig : 13_d}).

    \begin{table}
        \centering
        \begin{tabular}{ccc} \hline
            $\epsilon [^\circ]$ & $C_1^1 \leftrightarrow C_1^2$ & $\lambda$  \\ 
            \hline \hline 
            $ -10 $  & $(0.981,0,0.194) \leftrightarrow (0.211,0,0.977)$     & $0.408$   \\ 
            $ -11 $  & $(0.976,0,0.218) \leftrightarrow (0.237,0,0.972)$     & $0.454$   \\ 
            $ -12 $  & $(0.97,0,0.243) \leftrightarrow (0.265,0,0.964)$      & $0.502$   \\ 
            $ -13 $  & $(0.963,0,0.27) \leftrightarrow (0.293,0,0.956)$      & $0.551$   \\ 
            $ -14 $  & $(0.955,0,0.297) \leftrightarrow (0.324,0,0.946)$     & $0.601$   \\ 
            $ -15 $  & $(0.946,0,0.324) \leftrightarrow (0.356,0,0.934)$     & $0.652$   \\ 
            $ -16 $  & $(0.935,0,0.355) \leftrightarrow (0.391,0,0.92)$      & $0.703$   \\ 
            $ -17 $  & $(0.922,0,0.387) \leftrightarrow (0.428,0,0.904)$     & $0.756$   \\ 
            $ -18 $  & $(0.469,0,0.883) \leftrightarrow (0.906,0,0.423)$     & $0.809$   \\ 
            $ -19 $  & $(0.885,0,0.466) \leftrightarrow (0.515,0,0.857)$     & $0.863$   \\ 
            $ -20 $  & $(0.856,0,0.517) \leftrightarrow (0.57,0,0.822)$      & $0.918$   \\ 
            $ -21 $  & $(0.807,0,0.591) \leftrightarrow (0.645,0,0.764)$     & $0.974$   \\ \hline
            $ -22 $  & $(0.731,0,0.682) $                                    & $0.99$    \\ 
            $ -23 $  & $(0.729,0,0.685) $                                    & $0.971$   \\ 
            $ -24 $  & $(0.727,0,0.687) $                                    & $0.952$   \\ 
            $ -25 $  & $(0.725,0,0.689) $                                    & $0.935$   \\ \hline
        \end{tabular}
        \caption{The coordinates of attractive pure states in the Bloch sphere ($C_1^2 \leftrightarrow C_1^2$) for different values of the error parameter ($\epsilon$). The corresponding multipliers ($\lambda$) show that all of these cycles are attractive.}
        \label{table : pure attractive points 2}
    \end{table}
    
    Let us now examine the Julia set of the complex function $f_{\epsilon}$ describing the pure-state dynamics of the protocol in the presence of coherent errors. The Julia set in this case is constituted by the points of the border of the different basins of attraction. The change of its shape can reveal how the convergence pattern is changed when coherent errors are present. One can estimate the Julia set of $f_{\epsilon}$ by the iteration of the inverse of the complex function \cite{milnor2011dynamics}. We will refer to this procedure as "backward iteration". Since $f_{\epsilon}$ is a second-order complex rational function, it has two inverses
    \begin{equation}
        f_{\epsilon}^{-1}\left(z\right)=\pm\sqrt{\frac{\sin{\left(\frac{\epsilon}{2} + \frac{\pi}{4} \right)}- z \cos{\left(\frac{\epsilon}{2} + \frac{\pi}{4} \right)}}{\cos{\left(\frac{\epsilon}{2} + \frac{\pi}{4} \right)}+z \sin{\left(\frac{\epsilon}{2} + \frac{\pi}{4} \right)}}} \, ,
        \label{eq : inv_complex}
    \end{equation}
    meaning that every point has two preimages. In the simplest case, when one applies only one of the above inverse functions consecutively starting from a given point, then the resulting points  constitute two special branches of the backward iteration. 
    
    \begin{figure}
        \begin{center}
            \begin{subfigure}{0.45\columnwidth}
                \includegraphics[width=\linewidth]{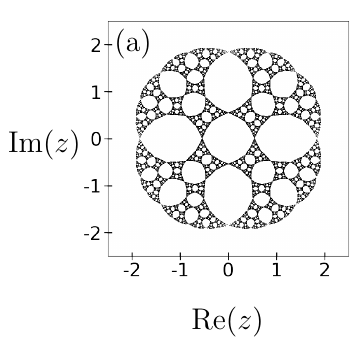}
                \phantomsubcaption
                \label{fig : 5_a}
            \end{subfigure}
            \begin{subfigure}{0.45\columnwidth}
                \includegraphics[width=\linewidth]{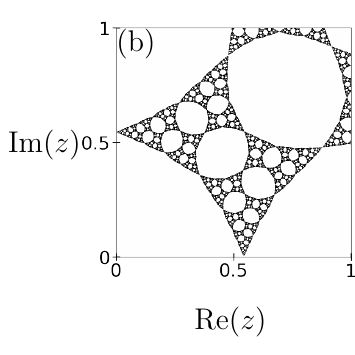}
                \phantomsubcaption
                \label{fig : 5_b}
            \end{subfigure}
            
            \begin{subfigure}{0.45\columnwidth}
                \includegraphics[width=\linewidth]{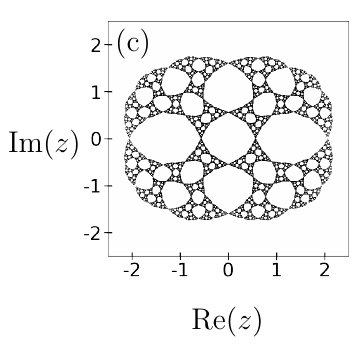}
                \phantomsubcaption
                \label{fig : 5_c}
            \end{subfigure}
                \begin{subfigure}{0.45\columnwidth}
                \includegraphics[width=\linewidth]{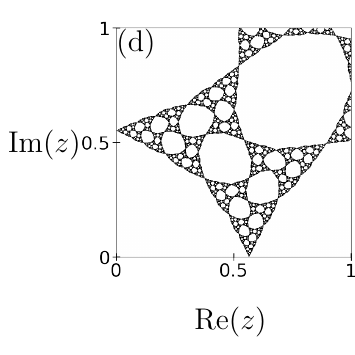}
                \phantomsubcaption
                \label{fig : 5_d}
            \end{subfigure}
            
            \begin{subfigure}{0.45\columnwidth}
                \includegraphics[width=\linewidth]{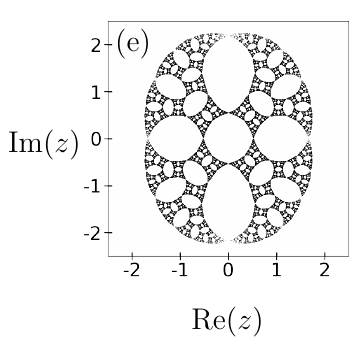}
                \phantomsubcaption
                \label{fig : 5_e}
            \end{subfigure}
            \begin{subfigure}{0.45\columnwidth}
                \includegraphics[width=\linewidth]{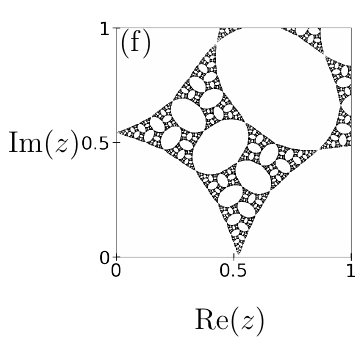}
                \phantomsubcaption
                \label{fig : 5_f}
            \end{subfigure}
            
            \caption{Julia sets of the $f_{\epsilon}\left(z\right)$ complex rational map for different values of  $\epsilon$. (a), (c), and (e) correspond to $\epsilon=0^\circ$, $4.5^\circ \left(5\% \right)$,  and $-4.5^\circ \left( -5\% \right)$, respectively. (b), (d), and (f) show the magnified structure of the Julia sets.}
            \label{fig : 5} 
        \end{center}
    \end{figure}
    
    In Fig.~\ref{fig : 5} we present points of the Julia set of different $f_{\epsilon}$ functions obtained by backward iteration (starting from one already known point of the Julia set and applying both inverses in every iteration). It can be seen that in the presence of coherent errors, the shape of the Julia set is deformed (cases \subref{fig : 5_c} and \subref{fig : 5_e}). For negative (positive) values of $\epsilon$, it contracts along the real (imaginary) axis and widens along the imaginary (real) axis (see Figs.~\ref{fig : 5_c} and ~\ref{fig : 5_e}, respectively). The deformation is not completely analogous for under- and overrotations (see also Figs.~\ref{fig : 5_b}, \ref{fig : 5_d} and \ref{fig : 5_f}). We also note here that according to our analysis, the distortion of the shape of the fractal occurs for smaller changes in $\epsilon$ in the case of underrotations, as compared to overrotations. 
    
\begin{figure}
        \begin{center}
            \begin{subfigure}{0.45\columnwidth}
                \includegraphics[width=\linewidth]{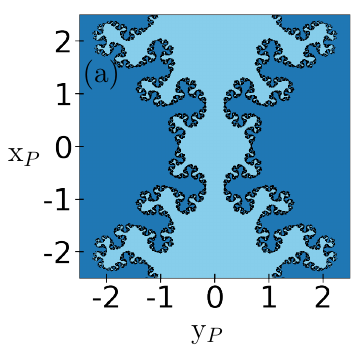}
                \phantomsubcaption
                \label{fig : 14_a}
            \end{subfigure}
            \begin{subfigure}{0.45\columnwidth}
                \includegraphics[width=\linewidth]{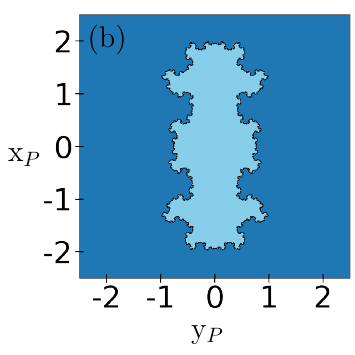}
                \phantomsubcaption
                \label{fig : 14_b}
            \end{subfigure}
            
            \begin{subfigure}{0.45\columnwidth}
                \includegraphics[width=\linewidth]{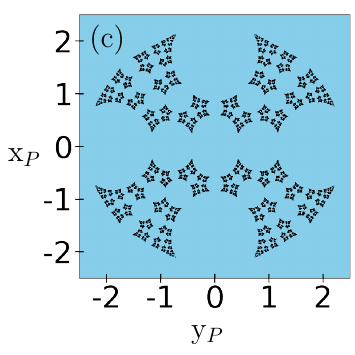}
                \phantomsubcaption
                \label{fig : 14_c}
            \end{subfigure}
            \begin{subfigure}{0.45\columnwidth}
                \includegraphics[width=\linewidth]{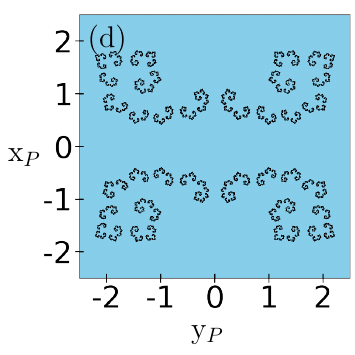}
                \phantomsubcaption
                \label{fig : 14_d}
            \end{subfigure}
            
            \caption{Stereographic projection of the surface of the Bloch sphere colored according to the convergence to the attractive points. Black points represent the Julia set. (a)-(d) figures correspond to $\epsilon=27^\circ \left(30\%\right)$, $45^\circ \left(50\%\right)$, $-27^\circ \left(-30\%\right)$ and $-45^\circ \left(-50\%\right)$, respectively.}
            \label{fig : 14}
        \end{center}
    \end{figure}

    Figures~\ref{fig : 14_a} and \ref{fig : 14_b} show the Julia set for two values of $\epsilon$ corresponding to over-rotations close to the limit where the maximally mixed state ceases to be attractive. It can be seen that although a stable pure length-2 cycle still exists (see also Figs.~\ref{fig : 13_a} and \ref{fig : 13_b}), the shape of the Juila set is very different from the ones seen in the presence of small over-rotations (see Fig.~\ref{fig : 5_e}). 
    In the case of an under-rotation corresponding to  $\epsilon \leq -21.5^\circ$ there is only one stable state on the surface of the Bloch sphere (see Figs.~\ref{fig : 13_c} and \ref{fig : 13_d}), therefore the structure of the Julia set is completely different: There is only one pure stable state (only one pure attraction region), thus the Julia set is disconnected (see Fig.~\ref{fig : 14_c}). Note that for even larger under-rotations, the Julia set remains disconnected for the same reason (see Figs. ~\ref{fig : 13_d} and \ref{fig : 14_d}).

\begin{figure} 
        \centering
        \includegraphics[width=0.9\columnwidth]{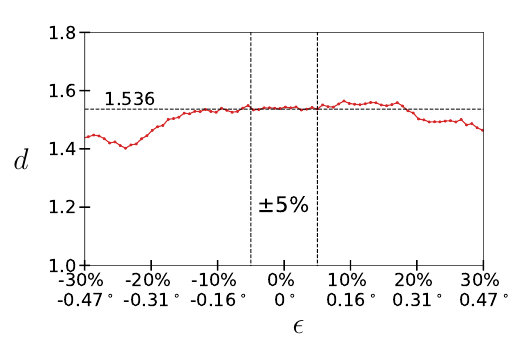}
        \caption{Numerically estimated fractal dimension $d$ of the Julia set of $f_{\epsilon}$ as a function of $\epsilon$.}
        \label{fig : 6} 
    \end{figure}

    The complexity of the Julia set is reflected by its fractal dimension, which can be estimated numerically by applying e.g. the box-counting method \cite{malachov2019phase}. In Fig.~\ref{fig : 6} we show the fractal dimension ($d$) as a function of the coherent error ($\epsilon$) which we obtained by applying a special version of the box-counting method introduced in \cite{mckay_efficient_2017-1}. One can see that the fractal dimension varies only to a small extent within the $\left[-5^\circ, 5^\circ \right]$ interval, corresponding to an under- or over-rotation of less than $6\%$ of the ideal rotation angle (i.e., $90^\circ$). In fact, its variation falls within the accuracy of the numerical computation.
    
     As we have seen above, the presence of a coherent error  changes both the possible final states and the pattern that separates the different convergence regions of initial states. The latter effect is more significant since due to the distortion of the very fine details of the border of attractive states, a given pure initial state might converge to the other attractive state and not the one that we would expect from the noise-free case. 
     
     \subsection{Dynamics of noisy initial states} \label{Dynamics of noisy initial states}
    
    In the coherent-error-free case we have seen that when state preparation errors are present, certain mixed initial states converge to the maximally mixed state, but the fractal nature of the border between the different attractive states is not necessarily destroyed (see Fig.~\ref{fig : 2_b}). Moreover, in Ref.~\cite{malachov2019phase} it was shown that the boundary of the basins of attraction of the two attractive pure periodic points -- which we will term  "pure attraction regions" -- on spherical surfaces corresponding to $P<1$ purities has a similar fractal structure as the Julia set, with the same fractal dimension. Therefore, we introduce here the term \textit{quasi-Julia set}, constituted by all mixed points of this dynamical system which are situated at the boundary of the pure attraction regions, being responsible for the fractal properties on lower purity surfaces, and possessing somewhat similar properties to the pure-case Julia set. 
    
    \begin{figure}[H]
        \begin{subfigure}{0.45\columnwidth}
            \includegraphics[width=\linewidth]{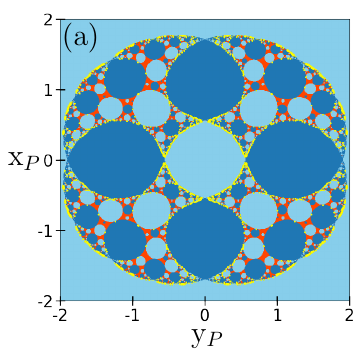}
            \phantomsubcaption
            \label{fig : 7_a}
        \end{subfigure}%
        \begin{subfigure}{0.45\columnwidth}
            \includegraphics[width=\linewidth]{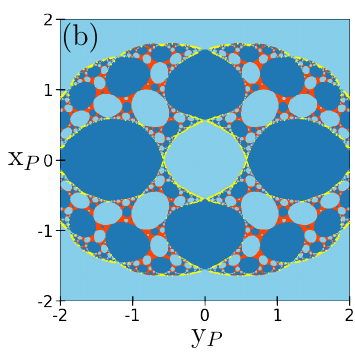}
            \phantomsubcaption
            \label{fig : 7_b}
        \end{subfigure}%
        
        \begin{subfigure}{0.45\columnwidth}
            \includegraphics[width=\linewidth]{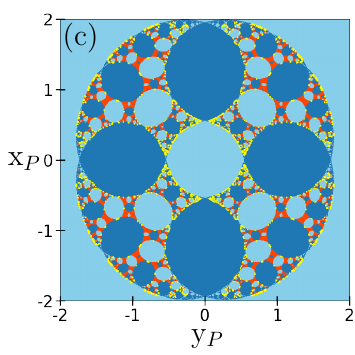}
            \phantomsubcaption
            \label{fig : 7_c}
        \end{subfigure}%
        \begin{subfigure}{0.45\columnwidth}
            \includegraphics[width=\linewidth]{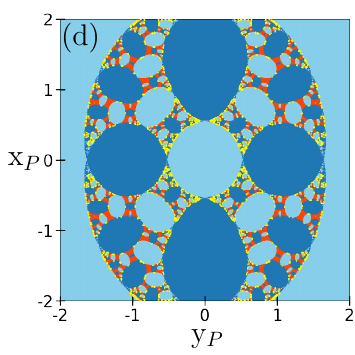}
            \phantomsubcaption
            \label{fig : 7_d}
        \end{subfigure}%
    
        \caption{Stereographic projections of the spherical surface corresponding to initial purity $P=0.95$ for different values of the coherent error. (a)-(d) figures correspond to $\epsilon=1.8^\circ \left( 2\%\right)$,  $4.5^\circ \left( 5\%\right)$, $-1.8^\circ \left(-2 \% \right)$, and $-4.5^\circ \left(-5\%\right)$, respectively. Light (dark) colour marks states that converge after an even (odd) number of steps to the first element of the attractive cycle $C_1$. Red colour indicates states that converge to the maximally mixed state ($C_0$). Yellow dots represent points of the quasi-Julia set, numerically calculated via backward iteration. }
        \label{fig : 7}
    \end{figure}
    
    Figure~\ref{fig : 7} shows stereographic projections of convergence regions on the spherical surface corresponding to initial purity $P=0.95$ for different values of the coherent error. One can see that the convergence regions, together with the quasi-Julia set, are distorted under the influence of coherent errors similarly to the Julia set (see Figs.~\ref{fig : 5_c} and \ref{fig : 5_e}). For under-rotations (Figs.~\ref{fig : 7_c} and \ref{fig : 7_d}), the quasi-Julia set widens along the $y$-axis and shrinks along the $x$-axis, while for over-rotation, the direction of the distortion is reversed (see Figs.\ref{fig : 7_a} and \ref{fig : 7_b}). The small structural details of the Julia set remain unchanged, suggesting that the information content of the convergence pattern is only slightly affected by the coherent error.
     
\begin{figure}[H]
        \begin{subfigure}{\columnwidth}
            \includegraphics[width=0.8\linewidth]{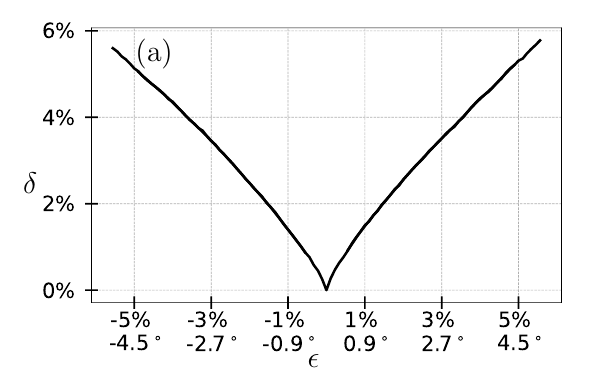}
            \phantomsubcaption
            \label{fig : 8_a}
        \end{subfigure}%
        
        \begin{subfigure}{\columnwidth}
            \includegraphics[width=0.8\linewidth]{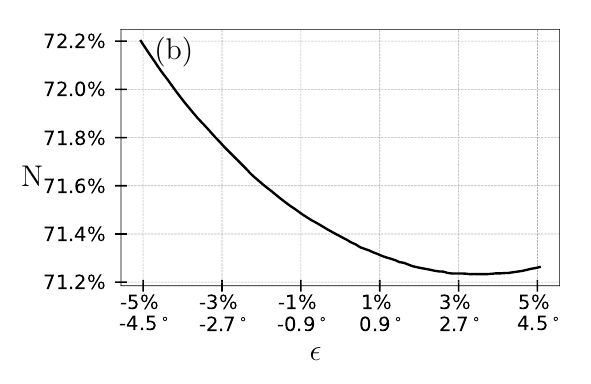}
            \phantomsubcaption
            \label{fig : 8_b}
        \end{subfigure}%
        
        \caption{(a) The ratio $\delta$ of uniformly chosen initial states that converge to a different attractor compared to the coherent-error-free case. A total of $10^6$ initial states were considered and the small displacement of the attractive states was neglected. (b) The percentage $N$ of uniformly chosen initial states that are purified by the protocol as a function of $\epsilon$.}
        \label{fig : 8}
    \end{figure}

     Because of the distortion of the convergence regions, however, a given noisy initial state which, without coherent error, would be purified by the protocol, may get maximally mixed instead. This represents a deviation caused by the faulty Hadamard gate. In order to quantify this effect, we have calculated the outcome of the protocol for uniformly chosen initial states from the entire state space and then determined the ratio of states that fall into the convergence range of a different attractive state compared to the coherent-error-free case (see Fig. \ref{fig : 8}). The results reveal that the protocol is sensitive to coherent errors. The system is disturbed already by small perturbations of the angle of rotation: In the case of an over- or under-rotation by $4.5^{\circ}$, about 5$\%$ of the noisy initial states converge to different attractive states compared to the coherent-noise-free case. We note here that in the calculations, we neglected the displacement  of the attractive states from their ideal values, as this is a much smaller effect than that of the distortion of the boundary of the convergence regions, which results in certain initial states converging to a different final state. Instead, we introduced an environment of radius $r$ around the original attractive states, where $r$ was equal to or greater than the displacement of the attractive periodic points due to the coherent error, and checked the convergence of the initial states with this tolerance. The calculation revealed that the main component of the deviation comes from the contribution of states that moved from one pure domain to another. The coherent error did not influence significantly the percentage of noisy initial states which get purified. In the case of under-rotation, this ratio is even slightly increased (see Fig.  \ref{fig : 8_b}).

    \subsection{Phase transition} \label{Phase transition in presence of coherent gate errors}
    
    The coherent-error-free protocol is robust to a certain degree of preparation noise (see Sec.~\ref{Characteristics of the ideal protocol}). The presence of coherent errors modifies the time evolution of the initial states, yet the primary features of the protocol remain similar. In what follows, we investigate how the preparation noise tolerance characteristics of the protocol are impacted when coherent errors are also present. 
    
\begin{figure}
        \begin{subfigure}{0.85\columnwidth}
            \includegraphics[width=0.8\linewidth]{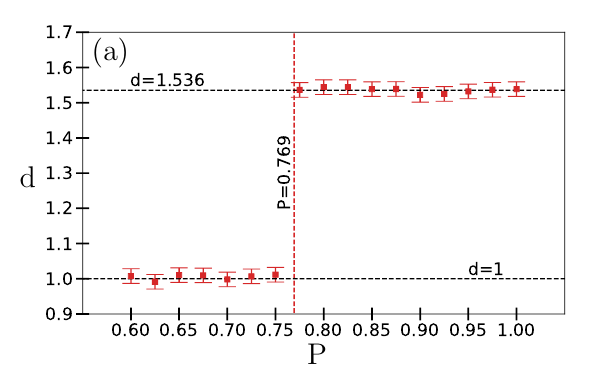}
            \phantomsubcaption
            \label{fig : 9_a}
        \end{subfigure}%
        
        \begin{subfigure}{0.85\columnwidth}
            \includegraphics[width=0.8\linewidth]{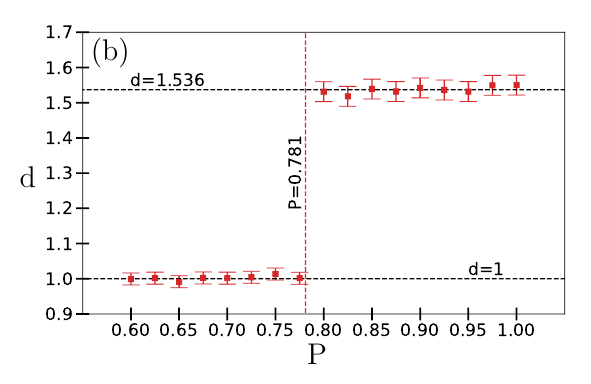}
            \phantomsubcaption
            \label{fig : 9_b}
        \end{subfigure}%
        
        \begin{subfigure}{0.85\columnwidth}
            \includegraphics[width=0.8\linewidth]{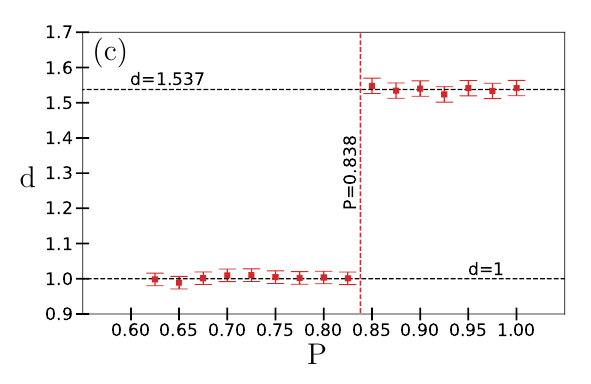}
            \phantomsubcaption
            \label{fig : 9_c}
        \end{subfigure}%
        
        \begin{subfigure}{0.85\columnwidth}
            \includegraphics[width=0.8\linewidth]{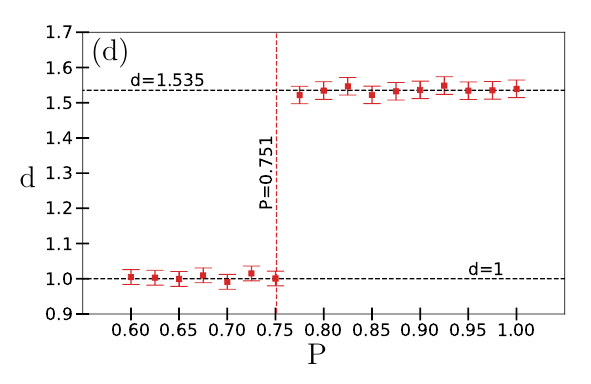}
            \phantomsubcaption
            \label{fig : 9_d}
        \end{subfigure}%
        
        \begin{subfigure}{0.85\columnwidth}
            \includegraphics[width=0.8\linewidth]{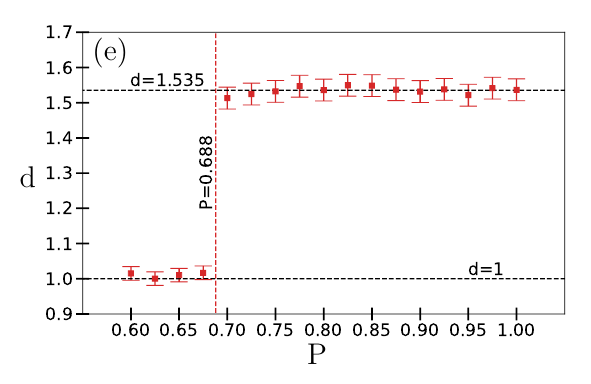}
            \phantomsubcaption
            \label{fig : 9_e}
        \end{subfigure}%
        
        \caption{Numerically estimated fractal dimensions of the boundary of different convergence regions as a function of the initial purity. (a)-(e) correspond to the $\epsilon=0^\circ$ (error-free), $0.9^\circ (1\%)$, $4.5^\circ (5\%)$, $-0.9^\circ (-1\%)$ and $-4.5^\circ (-5\%)$,  cases respectively. Horizontal dashed lines show the value of the fractal dimension of the corresponding Julia set. Vertical dashed lines mark the critical purity.}
        \label{fig : 9} 
    \end{figure}

    The preparation error tolerance of the coherent-error-free protocol was reflected by the fact that the fractal dimension of the convergence pattern remained constant up to a critical amount of noise (or equivalently, down to a critical purity $P_c$), where the disappearance of the fractal caused an abrupt change in the fractal dimension, similarly to a phase transition (see Fig.~\ref{fig : 9_a}). In Ref.~\cite{malachov2019phase} the authors observed that the fractal behaviour on $P<1$ purity surfaces can be attributed to the boundary of pure attraction regions, i.e., to the quasi Julia set. They also showed that the mixed repelling fixed point $C_3$ was the least pure state where the regions of attraction of the two pure states were in contact (see Fig.~\ref{fig : 2_d}). Above this noise level (below this purity), the two pure attraction regions had no common boundary, therefore, there was no fractal-like boundary section. According to the numerical results, the critical purity of the phase transition, which can be thought of as the threshold value above which the noise does not destroy the fractal, was precisely the purity of the mixed repelling fixed point $C_3$ located on the invariant plane at the common boundary point of the three different attraction regions.
    
    We have seen in Sec.~\ref{Properties of the invariant sets} that the presence of small coherent errors (i.e., $\epsilon\in[-10^{\circ},10^{\circ}]$) does not significantly change the important features of the protocol: There still exists an invariant plane in the dynamics, with an attractive pure length-2 cycle $C_1$ as well as a repelling mixed fixed point $C_3$. This suggests that the phase transition phenomenon might also survive the presence of such small coherent errors. Therefore, we performed a numerical analysis of the fractal dimension of the boundary of convergence regions as a function of the initial purity in the case of different values of $\epsilon$, as shown in Fig.~\ref{fig : 9}. We used the same method as in Sec.~\ref{Properties of the invariant sets} but this time on the stereographic projections of  spherical surfaces corresponding to different purities.  In Fig.~\ref{fig : 9}, one can see that the presence of a small coherent error does not lead to the disappearance of the phase transition phenomenon. The fractal dimension ($d$) remains constant (and equal to the dimension of the Julia set) down to some critical purity, where it suddently drops to $1$, indicating that the fractal disappeared. We directly estimated the purity where the fractal disappears (denoted by the vertical dashed lines in the plots). It can be seen  that the critical purity decreases for under-rotations (Figs.~\ref{fig : 9_d} and \ref{fig : 9_e}, while it increases for over-rotations (as shown in Figs.~\ref{fig : 9_b} and \ref{fig : 9_c}). The numerical findings also indicate that for under-rotations, the critical purity is equal to the purity of the mixed fixed point $C_3$, but if the coherent error is an over-rotation, this is no longer true, as the critical purity is lower than the purity of $C_3$ (this can be seen by comparing the values indicated in Figs.~\ref{fig : 9_d} and \ref{fig : 9_e} with those presented in Table~\ref{table : critical fixed point}). 
    
    The fact that for $\epsilon>0^{\circ}$ the critical purity is not the same as the purity of the mixed fixed point can be explained by examining Fig.~\ref{fig : 4} again. One can see that the presence of coherent errors deforms the attraction region of the maximally mixed state on the invariant plane: It is no longer symmetric to the axis $w$. As a result, the mixed repelling fixed point $C_3$,  located in the positive quadrant of the invariant plane, is actually not the lowest-purity touching point of the pure attraction regions as there are two such points (one in the third, one in the fourth quadrant) with lower purities (see Figs.~\ref{fig : 4_a} and \ref{fig : 4_b}). Since the fractal part of the boundary can be associated with the boundary of the pure attraction regions, one can expect that these points determine the critical purity of the phase transition.

    An effective method that was used in Refs.~\cite{malachov2019phase,portik_iterated_2022} to determine the critical purity was to iterate the inverses of the map of Eq.~(\ref{mixed_map}). There were two special branches of the backward iteration generated by applying iteratively only one of the inverses in each. In one of the branches, the sequence of preimages converged to the pure repelling fixed point $C_2$ situated at the intersection of the invariant plane and the surface of the Bloch sphere. In the other branch, the trajectories converged to the repelling mixed fixed point $C_3$, the purity of which was then identified as the critical purity. 
    
    In the case of coherent errors we can also determine the inverses of the map of Eq.~(\ref{eq : time_evoltion}), as presented in the Appendix, and identify these two branches of the backward iteration. Moreover, we find the exact same behavior: One of the branches purifies the points into $C_2$, while in the other branch, the trajectories converge to $C_3$, irrespective of the value of $\epsilon$. Consequently, this method cannot be used here to explain the fact that the critical purity in certain cases can be lower than the purity of $C_3$. 
    
    We have seen in Sec.~\ref{Properties of the invariant sets} that the Julia set, which coincides with the boundary of the pure attraction regions, can be identified through the repeated application of the inverses of the complex map, given in Eq.~(\ref{eq : inv_complex}). The preimages of any point of the Julia set are also elements of the set; therefore, during the backward iteration, the trajectory always remains on the boundary of the attraction regions. The numerical results suggest that, similarly to the case of the Julia set, the preimages of any point of the quasi-Julia set are also elements of the quasi-Julia set (see yellow points in Fig.~\ref{fig : 2_c} and Figs.~\ref{fig : 7}\subref{fig : 7_a}-\subref{fig : 7_d}). Hence, by the iteration of the inverses of the map of Eq.~(\ref{eq : time_evoltion}) presented in the Appendix, starting from an arbitrary point on the boundary of the pure attraction regions associated with a mixed state, it is possible to generate other quasi-Julia set points. 
    
    \begin{figure}
        \centering
        \includegraphics[width=\columnwidth]{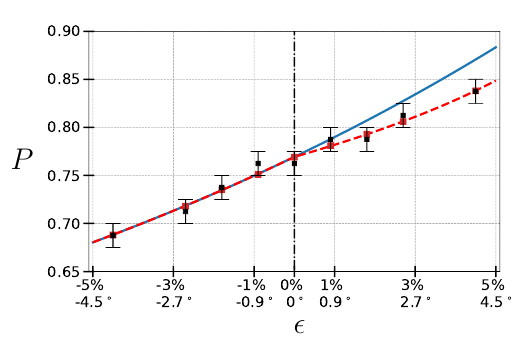}
        \caption{Critical purity of the phase transition as a function of $\epsilon$. The blue line shows the purity of $C_3$, while the red line shows the critical purity calculated from the preimages obtained by backward iteration. The black dots with error bars denote the values of the critical purity estimated numerically from the results of Fig.~\ref{fig : 9}.}
        \label{fig : 10} 
    \end{figure}
    
    In principle, one might be able to determine the entire quasi-Julia set by calculating all possible preimage sequences of a point from the set by evaluating all the combinations of the inverse functions. Unfortunately, the number of points grows exponentially with the number of iterations, therefore, this becomes infeasible already after a few steps. What one can do is to start from a given mixed point that is part of the quasi-Julia set and then determine its preimages by randomly choosing one of the two inverses in each iteration. This way, one can generate longer sequences of points, and the procedure can be repeated with other randomly chosen combinations of the inverses, to obtain a sufficient number of points from the quasi-Julia set. Obviously, this is not an effective method to get points on a given purity surface, but it can be a useful approach to see whether the pure attraction regions still have a common border below the purity of $C_3$ or not. Applying this method we have found that in the cases with $\epsilon>0^{\circ}$ there are indeed points of the quasi-Julia set, which have lower purities than that of $C_3$, but none of these were lower than the purities of the points in the third and fourth quadrant of the invariant plane, where the pure attraction regions still touch (see Figs.~\ref{fig : 4_c} and \ref{fig : 4_d}). In fact, these points are also preimages of $C_3$, with a given combination of the two inverses. Since no other combination of inverses produces a valid preimage of $C_3$, we conclude that the lowest purity quasi-Julia set points are one of these four points in the invariant plane ($C_3$ or its three preimages) and identify the critical purity of the phase transition with the purity of this lowest-purity point.

     The critical purities determined from the change of the fractal dimension and the purity of the preimages of $C_3$ on the invariant plane are in good agreement, as it can be seen in Fig.~\ref{fig : 10} and Table~\ref{table : critical purity}. 
    The difference between the two approaches is less than the accuracy of the calculation. Our findings numerically confirm that, although the emergence of coherent errors might alter the critical purity, the phase transition remains present. Additionally, let us point out that if the coherent error is an under-rotation (see Figs.\ref{fig : 9_d} and \ref{fig : 9_e}), then the critical purity is lower than in the coherent error-free case, enhancing, in a sense, the tolerance of the protocol against preparation noise.

    \begin{table} 
    \centering
    \begin{tabular}{ccccc} \hline
        $\epsilon [\%]$ & $\epsilon [^{\circ}$] & $P_3$ & $P_c$ & $P_c^{\text{est}}$ \\ 
        \hline \hline
        $ 5 $   & $ 4.5 $   & $ 0.871 $ & $ 0.838 $ & $0.8375 $ \\ 
        $ 3 $   & $ 2.7 $   & $ 0.827 $ & $ 0.806 $ & $0.8125 $ \\ 
        $ 2 $   & $ 1.8 $   & $ 0.807 $ & $ 0.793 $ & $0.7875 $ \\ 
        $ 1 $   & $ 0.9 $   & $ 0.788 $ & $ 0.781 $ & $0.7875 $ \\ 
        $ 0 $   & $ 0 $     & $ 0.769 $ & $ 0.769 $ & $0.7625 $ \\ 
        $ -1 $  & $ -0.9 $  & $ 0.752 $ & $ 0.751 $ & $0.7625 $ \\ 
        $ -2 $  & $ -1.8 $  & $ 0.735 $ & $ 0.735 $ & $0.7375 $ \\ 
        $ -3 $  & $ -2.7 $  & $ 0.718 $ & $ 0.718 $ & $0.7125 $ \\ 
        $ -5 $  & $ -4.5 $  & $ 0.688 $ & $ 0.688 $ & $0.6875 $ \\ \hline
    \end{tabular}
    \caption{Purity $P_3$ of $C_3$, the critical purity $P_c$ from the lowest-purity quasi-Julia set point calculated via backward iteration, and the estimated critical purity $\left( P_{c}^{\text{est}} \right)$ from the direct calculation of where the fractal vanishes, for different values of the coherent error parameter $\epsilon$. The uncertainty of the directly estimated critical purity is equal to $0.0125$, which is the resolution of the numerical calculation.}
    \label{table : critical purity}
    \end{table}
    
    \subsection{Small coherent error} \label{Small coherent error}
    
    We have seen in the previous sections that for small coherent errors, the main characteristics of the protocol and the phase transition phenomenon are not altered significantly. Based on the results, we determine the maximum magnitude of the coherent error for which the protocol remains similar to the coherent-error-free case in the sense that there still exists an attractive pure length-2 cycle into which noisy initial states may get purified, as well as a repelling mixed fixed point, which can be associated with a phase transition critical point. 
    
    For coherent errors representing under-rotations with a magnitude larger than $21.5^\circ$, the structure of the dynamics is significantly altered since there is no attractive length-2 cycle on the surface of the Bloch sphere. All states which  purify converge to a single attractive pure fixed point. In the case of over-rotations, the most rapidly changing characteristic of the protocol is the purity of the internal fixed point $C_{3}$, thus, the critical point of the phase transition. For over-rotations larger than $10^\circ$ there is no internal fixed point.
    
    However, for small coherent errors in the $\left[-10^\circ,10^\circ\right]$ symmetric error range, all essential properties of the protocol change only slightly: The structure of the corresponding basins of attraction ranges is distorted, but all the phenomena experienced in the error-free case take place in a similar way. Although the critical purity of the phase transition significantly varies, the phase transition still takes place. Knowing that the current quantum processors have an error rate of less than $1\%$ for single- and two-qubit operations \cite{Preskill2018quantumcomputingin, 00050}, we can state that the studied nonlinear quantum protocol is robust against coherent gate errors.
    
    \section{Discussion} \label{Discussion}
    
    We studied the effect of coherent quantum logic gate errors on a specific iterated quadratic nonlinear quantum protocol. We assumed a scenario, where the coherent error affects the Hadamard gate applied in every step of the protocol. Since the Hadamard gate is usually implemented as a sequence of $Z$ and $X$ gates, where the $Z$ gates are only virtual, we trace back the error to a miscalibrated single-qubit $X$ gate, and described the coherent error as an over- or under-rotation. We determined the relations describing the evolution of an arbitrary (pure or mixed) qubit state and examined the effects of the occurring coherent gate errors. We have shown that for small coherent errors, the characteristic properties of the protocol are slightly distorted: in the preparation noise-free case the attractive pure states are displaced, and the border of their convergence regions are deformed, but the fractal nature of this border remains. In addition its fractal dimension is not significantly changed. However, as a result of the distortion of the delicate fractal pattern, a given pure initial state might converge to a different pure state than in the ideal protocol. In the case when preparation noise is also present, we showed that all relevant features of the coherent-error-free case survive, namely, the invariant plane still exists and contains the relevant fixed points and cycles of the dynamics, which are slightly shifted, and the border of the convergence regions continues to be a fractal, though, similarly to the pure case, somewhat deformed. We pointed out that, if the initial state is chosen from regions less affected by the distortion, then the adverse effects of coherent errors can be eliminated. Hence, we are able to identify regions of reliable operation.
    
    We investigated the most notable characteristic of the original protocol, the preparation noise tolerance property manifested as a phase transition of the fractal dimension of the the border of convergence regions as a function of the initial purity. We have shown that the presence of small coherent errors does not destroy this property, although the critical purity of the phase transition shifts from its original value. Remarkably, we found that as a result of the distortion of the convergence regions on the invariant plane of the dynamics, the critical purity cannot always be identified with the same type of repelling point as in the coherent-error-free case, but rather with its lowest purity preimage on the invariant plane.
     
     We also investigated how large coherent errors affect the dynamics and identified thresholds of the coherent error for both over- and under-rotations, where the above mentioned similarities with the original dynamics are completely lost. 
    
    Our work demonstrates that despite the fact that the outcome of a quantum protocol can be affected by coherent errors, as long as the coherent error is small,  the characteristic properties of the time evolution remain unchanged. In this respect, our general finding suggests a certain resilience of the fragile chaotic dynamics against noise and errors.
    This fact may encourage the utilization of such nonlinear protocols and the search for new applications in addition to the existing ones. Nevertheless, further examination of the delicate components of the system is necessary in order to identify all potential sources of error if the protocol is to be utilized in a practical setting. As the key ingredients of these protocols are the measurement and the post-selection steps, it is extremely important to examine the consequences of their errors on the dynamics in these schemes. 
    
    \section{Acknowledgement}
    
    We acknowledge support from the National Research, Development and Innovation Office of Hungary, project No. TKP-2021-NVA-04 and support from OpenSuperQPlus100. Igor Jex acknowledges the financial support from the state budget under RVO14000 and the Grant Agency of the Czech republic (GAČR) under Grant No. 23-07169.

    \appendix*
    \section{Inverse map}
    
    Backward iteration makes it possible to calculate the points of the quasi-Julia set and the preimages of the internal repelling fixed point, one of which is the critical point of the phase transition. This procedure requires the iterative application of the inverse time evolution maps. In order to determine the inverses of Eq.~(\ref{mixed_map}) , one can rearrange Eq.~(\ref{transf_rho}) as
    \begin{equation}
    \tilde{H}^{-1}\rho^{(n+1)} \left(\tilde{H}^{\dagger}\right)^{-1} = \dfrac{\rho^{(n)} \odot \rho^{(n)} }{\Tr \left( \rho^{(n)} \odot \rho^{(n)} \right)} 
    \end{equation}
    where, $\tilde{H}$ is the faulty Hadamard gate introduced in  Eq.~(\ref{Eq : 5}). $\tilde{H}$ is unitary, which results in that $\tilde{H} = \tilde{H}^{-1} = \tilde{H}^\dagger$, therefore one can rewrite the equation as
    \begin{equation}
    \tilde{H}\rho^{(n+1)} \tilde{H} = \dfrac{\rho^{(n)} \odot \rho^{(n)} }{\Tr \left( \rho^{(n)} \odot \rho^{(n)} \right)} \, .
    \label{Eq : 10}
    \end{equation}
    We parameterize the density operator of the initial state with lowercase letters and the density operator of the final state with uppercase letters corresponding to the coordinates of the Bloch vector. The left side of the equation can be easily derived 
    \begin{equation}
    \left[ 
    \begin{array}{lr}
    1 - \sin{\left( \epsilon \right)} U + \cos{\left( \epsilon \right)} W 
    & \cos{\left( \epsilon \right)} U + \sin{\left( \epsilon \right)} W - i V \\
    \cos{\left( \epsilon \right)} U + \sin{\left( \epsilon \right)} W + i V
    & 1 + \sin{\left( \epsilon \right)} U - \cos{\left( \epsilon \right)} W
    \end{array}
    \right] \, , \notag
    \end{equation}
    at the same time, according to the definition, the right side is
    \begin{equation}
    \dfrac{1}{2\left(1+w^2\right)}
    \left[
    \begin{array}{lr}
    \left(1+w\right)^2 & \left(u-iv\right)^2 \\
    \left(u+iv\right)^2 & \left(1-w\right)^2 
    \end{array}
    \right] \,. \notag
    \end{equation}
    From the diagonal part of the matrix equation, one can express a combination of the $w$ coordinate
    \begin{equation}
    \left(\dfrac{1-w}{1+w}\right)^2 = \dfrac{1 + \sin{\left( \epsilon \right)} U - \cos{\left( \epsilon \right)} W}{1 - \sin{\left( \epsilon \right)} U + \cos{\left( \epsilon \right)} W} \, ,
    \end{equation}
    from this equation, one can express $w$ as
    \begin{equation}
    w = \frac{1 - \sqrt{\dfrac{1 + \sin{\left( \epsilon \right)} U - \cos{\left( \epsilon \right)} W}{1 - \sin{\left( \epsilon \right)} U + \cos{\left( \epsilon \right)} W}}}{1 + \sqrt{\dfrac{1 + \sin{\left( \epsilon \right)} U - \cos{\left( \epsilon \right)} W}{1 - \sin{\left( \epsilon \right)} U + \cos{\left( \epsilon \right)} W}}}  
    \end{equation}
    Based on the derived value of $w$ and the off-diagonal part of Eq.~(\ref{Eq : 10}), one can express the other two coordinates with the Bloch coordinates of the image state as
    \begin{align}
    u =  \text{Re}\left( \sqrt{2\left(1+w^2\right) \left[\cos{\left( \epsilon \right)} U + \sin{\left( \epsilon \right)} W + i V \right]} \right)  \, , \\ \notag
    v =  \text{Im}\left( \sqrt{2\left(1+w^2\right) \left[\cos{\left( \epsilon \right)} U + \sin{\left( \epsilon \right)} W + i V \right]} \right)\, .
    \end{align}
    
    \bibliography{MyBIB.bib}

\end{document}